\documentclass{revtex4-2}
\usepackage{graphicx}% Include figure files
\usepackage{dcolumn}% Align table columns on decimal point
\usepackage{bm}% bold math
\usepackage{gensymb}
\usepackage{amsmath}
\usepackage{float}
\usepackage{xcolor}
\usepackage{natbib}
\usepackage{soul}

\begin{document}

\preprint{APS/123-QED}
\title{Patterning surface textured plates with a viscoplastic fluid}
\author{Vanessa R. Kern}
\affiliation{Mechanics Division, Department of Mathematics, University of Oslo, 0316 Oslo, Norway}
\author{Marcel Moura}
\affiliation{PoreLab, the Njord Center, Department of Physics, University of Oslo, 
NO--0316 Oslo, Norway}
\author{P{\aa}l E. S. Olsen}
\affiliation{Mechanics Division, Department of Mathematics, University of Oslo, 0316 Oslo, Norway}
\author{Andreas Carlson}
\email{acarlson@math.uio.no}
\affiliation{Mechanics Division, Department of Mathematics, University of Oslo, 0316 Oslo, Norway}
\affiliation{Ume{\aa} University
Department of Medical Biochemistry and Biophysics, 901 87 Ume{\aa}, Sweden}

\date{\today}

\begin{abstract}
The deposition of a viscoplastic fluid onto a substrate can be achieved by simply moving apart two plates initially separated by a fluid filled gap, where the footprint shape depends on the initiation of a fingering instability. Here, we present another approach for the controlled deposition of a viscoplastic fluid by designing the macroscopic structures of the solid substrate. Through experiments in a lifted Hele-Shaw cell, we explore how slot, square, pyramid, and triangle-shaped patterns affect the dynamics of liquid deposition. These substrate structures directly control the final shape of the viscoplastic fluid footprint. It turns out that these substrate patterns have little influence on the normal adhesive force acting on the plates.
\end{abstract}
\maketitle

\section{Introduction}
Deposition of a complex fluid onto a substrate appears across a wide range of applications and natural processes, including how saliva coat tongues \cite{noel2017frogs}, the spreading mayonnaise on bread \cite{boyaci2024transition}, and is also found in medical diagnostics \cite{robinson2002microarray}. One such type of non-Newtonian fluid is a yield stress or viscoplastic fluid, which can sustain a stress without flowing, i.e., they behave as solids until a critical stress threshold is exceeded \cite{COUSSOT201431,nguyen1992,bonn2017}. These fluids are found in industrial applications (drilling fluids \cite{BlandineSoftMatter}), in food processing (mayonnaise, ketchup, chocolate \cite{garg2021fluidisation}), in cosmetics (tooth-paste, hair gels), as well as in geological lava flows \cite{TS2023,NeufeldMountain}. Yield stress fluids are topical in the development of 3D printing \cite{geffrault2023printing, colanges20232}, which rely on the detailed control of the deposition of the fluid on a solid substrate. We demonstrate here a route to controlling the deposition pattern of yield stress fluids through the design of  macroscopic substrate textures when simply separating two plates.

When a viscous fluid enters a narrow gap filled with a more viscous immiscible fluid, the interface can become unstable and starts forming a finger-like pattern. The transition from a circular advancing interface to a finger-like pattern has been well characterized for Newtonian liquids \cite{paterson1981,chen1989,miranda1998}, which is governed by the viscosity contrast, the surface tension and the flow rate, and known as the \textit{Saffman–Taylor instability} \cite{SaffmanTaylor,Saffman_1986}. If the liquids are viscoplastic the interface pattern changes \citep{Irm020,bonn2017,LindnerPRL} and the description of the stability threshold becomes more involved \cite{COUSSOT_1999,fontana2013}. A yield stress fluid profoundly alters the fingering dynamics \cite{LindnerPRL,COUSSOT_1999,ball2021} and as such affect deposition of liquid on a surface using a lifted Hele-Shaw cell \citep{Irm020}.

Experiments have shown a criterion for the onset of fingering in colloidal gels under tension \citep{Irm020}, distinguishing a stable (elastic deformation) from an unstable (air invasion and fingering) regime. It has been argued that a fingering instability can only be initiated if a sufficiently large stress is locally transferred to fluidize the gel at the finger tips. Once the fluid yields, the patterns that form appear to have strong similarities with the classic viscous-fingering scaling predictions for Newtonian fluids, but where the difference stems from parts of the liquid appearing to be in yielded and unyielded state. In addition to the yield stress properties of the fluid, also the interaction between the fluid and the confining surfaces can affect the flow pattern \cite{TalalNatPhys, Amy2022}. Model yield stress fluids as Carbopol can also exhibit fluid slip if the surface is smooth \cite{DANESHI201965,behbood2022}. One way to ensure a no-slip condition on the surface is to introduce surface texture or to chemically functionalize the material \cite{Jalaal2015}. It has been demonstrated that wall slip can indeed influence finger growth in viscoplastic fluids \citep{dufresne2023}.

In the context of adhesion, Bikerman’s pioneering work \citep{bikerman1947} established that the normal force required to separate two plates joined by a fluid depends both on surface energy and bulk deformation (and therefore in the adhesive's rheology). It was later shown that yield stress fluids sustain a significant force even when fingering occurs \citep{derks2003} due to the resistance of the bulk material to cohesive failure and the surprising result that the fingering instability did not seem to have any significant effect on the work of adhesion. These works paint the picture that the interplay between substrate roughness (slip lines and pinning sites), and complex rheology determines not only the final pattern morphology of the process, but also the forces imposed by the fluid on the surrounding solid boundaries, with immediate impact on adhesion applications. In this work we explore how a surface-textured plate affects the flow and final deposited footprint of a yield stress fluid during the separation of two circular plates in a Hele-Shaw cell. By varying texture geometry, initial plate separation and pull-off speed, we identify and classify the interfacial patterns, as well as establishing its effects on the adhesive (normal) force on the lifted plates.

\begin{figure}[t!]
    \includegraphics[width=1.0\columnwidth]{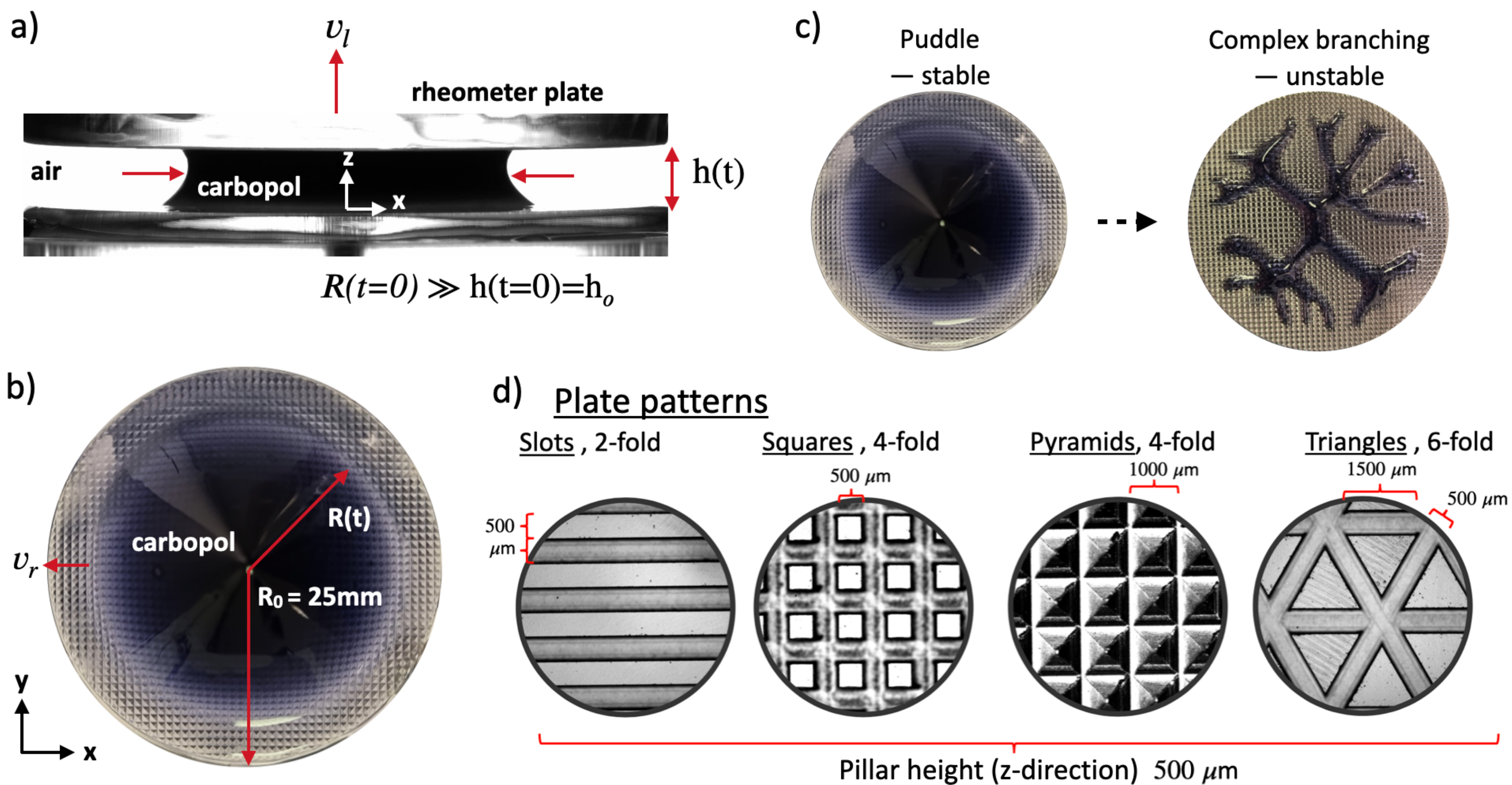}
    \caption{a) Side-view image of the experimental setup, where a Carbopol mixture is used as a model yield stress fluid and squeezed between two circular plates in a rheometer. Initially ($t=0$) the plates are separated by a distance $h_0$ that is varied in the experiments (around one millimeter). As the plates are lifted by a velocity $v_l$ air enters into the gap, forming in the end a single bump (puddle) or a complex pattern. b) Image of the bottom plate gives way to illustrate the deposited pattern and its radial dynamics $R(t)$, where liquid initially fills the gap and entirely wets the plate with $R_0=25$mm. c) Two characteristic features observed in the experiments, where the yield stress fluid either leaves a single bump on the plate or leaves a finger-like structures on the substrate. d) The influence of substrate pattern has been explored through different shapes including parallel slots, squares, pyramids and triangles, where all these shapes have a height of $500\mu$m. These micro-milled surfaces are also semi-transparent, still allowing for bottom and side view imaging.}

  \label{fig:Fig1}
\end{figure}
\section{Experimental setup} 
To test how the deposition pattern of a yield stress fluid can be controlled by the substrate topography/pattern, we designed an experimental setup around an Anton Paar {702} rheometer, see Fig. 1a. By lifting the upper plate, the rheometer is equivalent to a lifted circular Hele-Shaw cell. The use of the rheometer allows precise control of the initial distance $h(t=0)=h_0\in[0.1-5]$mm that is separating the two plates in the wall-normal direction, where initially $t=0$ we always have a fluid fully wetting the rheometer plates with radius $R_0=25$mm. %For most of our experiments, the condition $R_0/h_0\ll 1$ is met.
In addition, the rheometer provides control of the lifting speed of the upper plate $v_l$ and we can simultaneously measure the plates wall-normal force $F_N(t)$ with time. The dynamic interface of the yield stress fluid is monitored through side and bottom view imaging, allowing for characterisation of the footprint it leaves on the bottom plate, giving a measure of the effective wetted radius $R(t)$ and the radial speed of the liquid front $v_r$ (Fig. 1b). To enable bottom view imaging, the plates were made from transparent Plexiglas. A tilted mirror positioned beneath the lower plate reflects the image toward a camera mounted on the side of the setup.

After the yield stress fluid is sandwiched between the rheometer plates, the upper plate is lifted with a constant speed $v_l\in[10-1000]\mu$m/s. Consequently, the gap between the plates increases linearly with time $h(t)$ and the yield stress fluid is pulled to the center of the plates as air enters the gap. In this process, the wetted area on the plates reduces as air is pulled in and the intrusion dynamics can transition from the formation of an axially symmetric puddle with a nearly constant radial interfacial velocity $v_r$ in the azimuthal direction to a highly complex (unstable) branching pattern, see Fig. 1c, as also demonstrated by others \cite{Irm020,derks2003,dufresne2023}. Our interest lies in understanding how the dynamic retraction pattern and the final deposited footprint are affected by a macroscopic surface pattern of the two plates. Fig. 1d illustrates the design of the different semi-transparent plates with structures of slots, pillars, pyramids, triangles, which are used to quantify the effect of the patterning on the adhesive (normal) force and the deposited footprint of a yield stress fluid represented here by a Carbopol mixture.

The Carbopol samples were created by combining Carbopol powder with 1M NaOH in deionized water (dyed with Nigrosin), followed by mechanical mixing of the solution for about 10 days. The Carbopol to deionized water ratio was varied, while the ratio of added NaOH to Carbopol powder was around 8 across all samples. The characterization of Carbopol solutions was conducted on an Anton-Paar Rheometer 702 using striated parallel plate geometry to minimize slip, and this was done prior to each experimental set. A pre-shear step was incorporated before each measurement to effectively disrupt and then reform the fluid's microstructure, see \cite{kern2022} for further details about the preparation procedure. The different Carbopol samples where fitted to the Herschel-Bulkley model, leading to an yield stress value of $\tau=30$Pa. One caveat with using Carbopol is that it is well known to produce slip on smooth surfaces. All the results reported in this article on smooth surfaces are with a surface treatment following a protocol similar to \citep{Jalaal2015,dufresne2023,kern2022}. Note that the structured surfaces have millimeter sized features that would naturally reduce any slip effects. 

\begin{figure}[t!]
        \includegraphics[width=0.9\columnwidth]{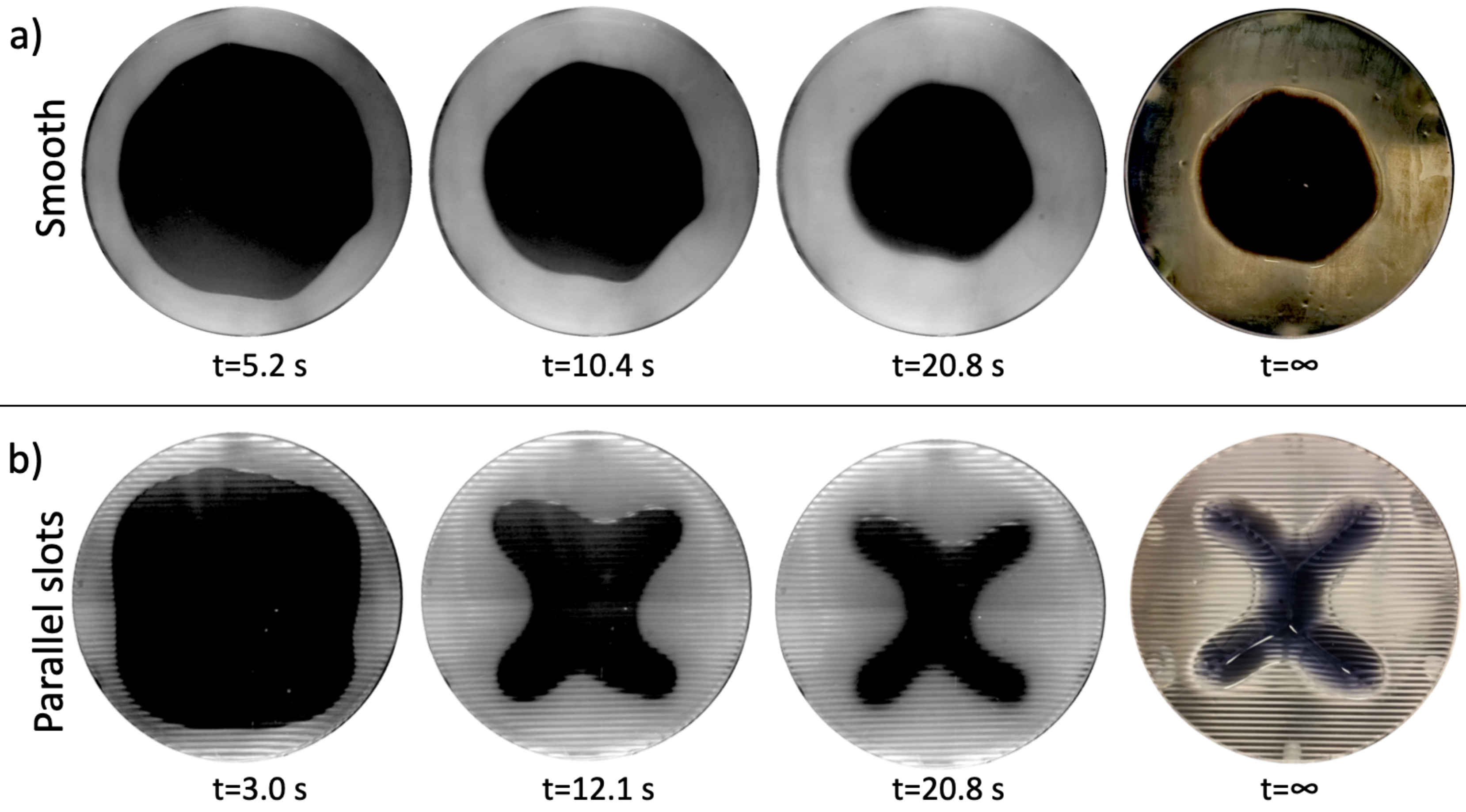}
    \caption{Bottom-view images illustrate the interfacial dynamics as two plates in a Hele-Shaw geometry are separated at a velocity of $v_l = 160\,\mu\text{m/s}$. The effective mean initial height is $h_{0,eff} = 1\,\text{mm}$, and the fluid is a Carbopol mixture with a yield stress of $\tau = 30\,\text{Pa}$. Panel a) shows the case with smooth plates (treated to enforce a no-slip boundary condition) while panel b) corresponds to plates patterned with parallel slots. In the slotted case, the interface shape evolves as the yield stress fluid retracts radially producing a characteristic cross-shaped deposition pattern. In contrast, the smooth plates generate a more symmetric invasion front, leaving a roughly circular bump. To account for the fluid volume retained within the slots, the initial separation in the slotted case is reduced to $h_0 = 0.75\,\text{mm}$, ensuring the same mean film thickness as in the smooth configuration ($h_{0,eff} = 1\,\text{mm}$), as the depth of the slots is $0.5\,\text{mm}$, see Fig. 1d. The rightmost images in both a) and b) are taken from above after the plates are disconnected by the fluid.}
  \label{fig:Fig2}
\end{figure}

By using an effective viscosity on the order of $\mu \sim 10 \ \text{Pa}\cdot\text{s}$, the flow is characterized by low Reynolds and moderate Capillary numbers. The characteristic shear rate is estimated as $\dot{\gamma} = v_r/h_{0,eff}$, where $v_r$ is the initial radial velocity and $h_{0,eff}$ is the initial mean fluid thickness. The quantity $h_{0,eff}$ is introduced to accomodate for the fluid volume that is also found in the grooves of the plate surface patterns (additional details in the coming section). The Reynolds number is estimated as $\text{Re} = \rho v_r h_{0,eff}/\mu$,
and the Capillary number as $\text{Ca} = \mu v_r/\gamma$, where $\rho \approx 1000 \ \text{kg/m}^3$ is the fluid density and $\gamma \approx 0.05 \ \text{N/m}$ is the surface tension \cite{kern2022}. For the experimental conditions reported here, we find that for most experiments $\text{Re} \in [10^{-5}, 10^{-3}]$, indicating negligible inertial effects, while $\text{Ca} \in [0.1, 1]$, suggesting that viscous forces are comparable to or dominate over capillary forces during the dynamics.

\begin{figure}[t!]
    \includegraphics[width=1.0\columnwidth]{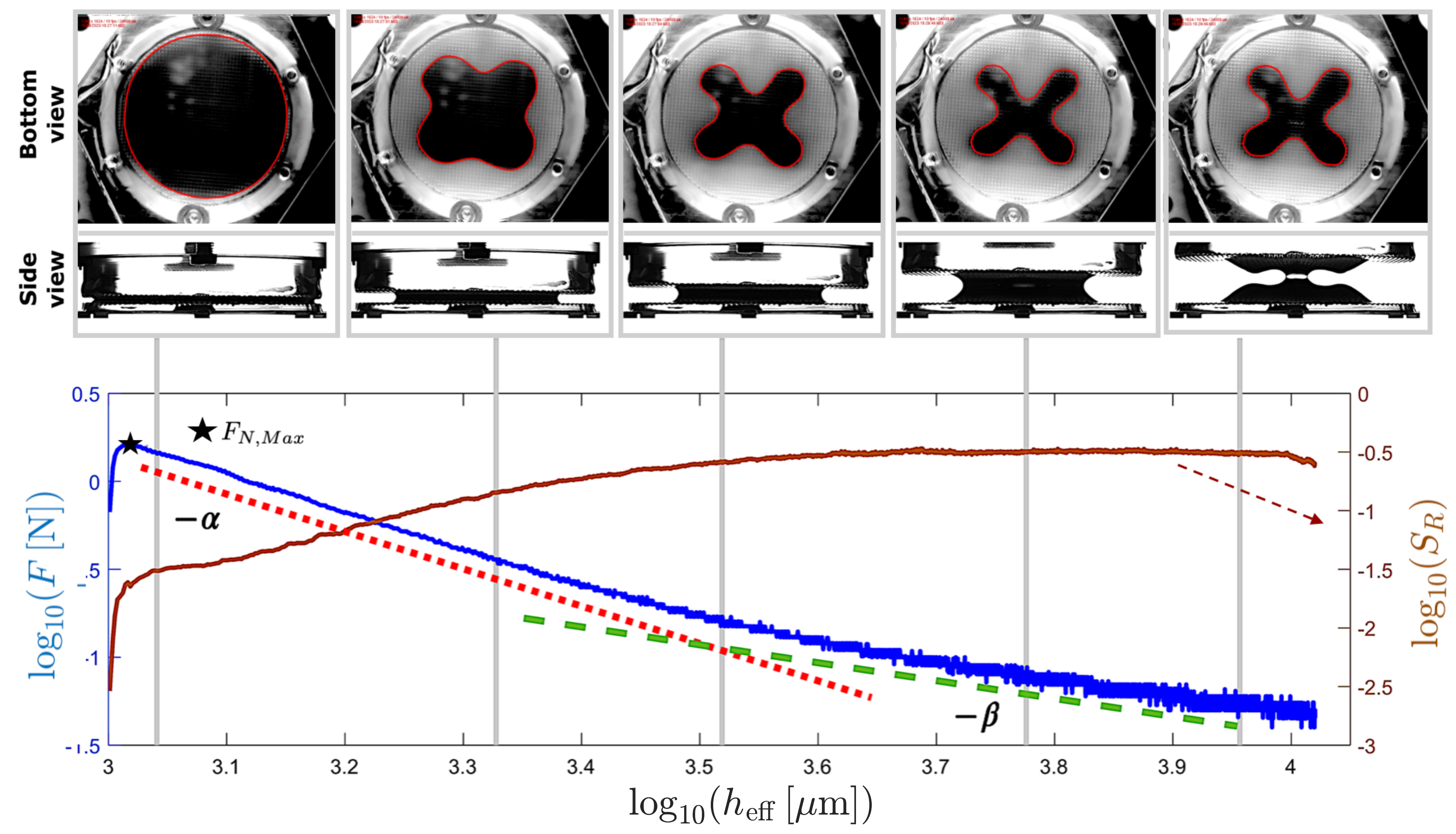}
 \caption{Top row: Panels showing the bottom and side view image of the yield stress fluid interface The radius of the plate $R_0=25$mm acts as a scale bar. Bottom plot: Variation of the pulling force (blue, left axis) and the geometric factor $S_R$ (brown, right axis) as a function of the effective cell gap $h_{eff}$ for an experiment using pyramid-patterned plates. The pulling speed and initial effective gap were $v_l = 51.2\,\mu\text{m/s}$, $h_{0,eff} = 1\,\text{mm}$ and $\tau=30$Pa, respectively. Two lines are added as a guide-to-the-eye lines, where the dotted line indicate the slope $\alpha$ and the dashed line the slope $\beta$ describing the force as a function of gap height. The bottom and side view images of the interface are collected in five panels placed above the plot with the corresponding time points illustrated by the vertical solid gray lines. The crossover between the force regimes described by $\alpha$ and $\beta$ coincides with the stabilization of the geometric factor $S_R$. As the gap height extends beyond this point, the dynamics is primarily driven by an extensional flow and neck thinning (eventually leading to structural breakup) as the contact line is pinned. A video of the experiment and measurements shown here can be found in the Supplemental Material.}
\label{fig:Force_Sr}
 
\end{figure}

\section{Results}

To illustrate how surface texture affect the intrusion dynamics and the patterning between the two plates, we show in Fig. 2 the results for a smooth (upper row) and patterned plate (bottom row) for the same lifting velocity $v_l=160\,\mu\text{m/s}$ and mean film thickness $h_{0,eff} = 1\text{mm}$. When comparing the footprint shapes at different time points, it is clear that the surface texture affects the dynamics. On the smooth plate, we notice that the footprint shrinks radially forming a circle that essentially ``shrinks'' with time and leaves a fairly symmetric puddle. It has been suggested that as long as the energy input to the system goes beyond a threshold value, the fingering instability  would be independent of the fluid's yield stress \cite{Irm020}. As we see in Fig. 2a the experiments are in the stable regime and no air fingers form. However, if we now keep the same fluid, effective film thickness and lifting speed but make a substrate pattern consisting of macroscopic parallel slots features, we notice a clear difference in the dynamics as compared to the smooth plate, as seen in Fig. 2b. Instead of a symmetric puddle, a cross-shaped footprint of the yield stress fluid is deposited. It is clear that the four finger-like structures are much larger than the substrate pattern, where the radial velocity towards the center of the plate is highest around the ``equator'' and along the ``prime meridian''. Thus, it is clear that the pattern on the plates affects the final footprint of the deposited yield stress fluid.

As the plates separate, the normal adhesive force will change with time. In Fig. 3 we illustrate how the generated force correlates with the footprint of the yield stress fluid. At first, air intrudes uniformly into the gap and the normal force quickly increases with a peak that sets the maximal normal force $F_{N,max}$ as illustrated by the star marker (Fig. 3a). From the point of $F_{N,max}$ the force monotonically decreases following a power-law $\sim h_{eff}^{-\alpha}$. As the force decays with increasing gap $h$ the system reaches a crossover region after which the slope of the curve changes to $\sim h_{eff}^{-\beta}$, as also observed by \cite{Irm020}. Our experiments enabled simultaneous visualization of the evolving pattern from both the side and the transparent bottom of the cell, while also recording the corresponding force measurements. Figure 3 presents selected snapshots taken at the gap heights indicated by the vertical gray lines. The first panel to the left shows at early times the initially circular invasion front, with its perimeter highlighted with a red line. As time progresses, a cross-shaped pattern begins to emerge, as shown in the second panel on the left. The center panel marks the approximate moment when the pattern at the bottom of the cell appears to freeze, i.e., the contact line becomes pinned, and the flow is primarily extensional, with minimal change in the wetted area on the plate, as illustrated in the fourth panel from the left. The rightmost panel captures the moment just before the complete separation of the top and bottom fluid regions, characterized by the thinning of the neck observed in the side-view imaging. A video showing the temporal evolution of the fluid pattern and the associated force measurements can be found in the Supplemental Material.

\begin{figure}[b!]
    \includegraphics[width=1.0\columnwidth]{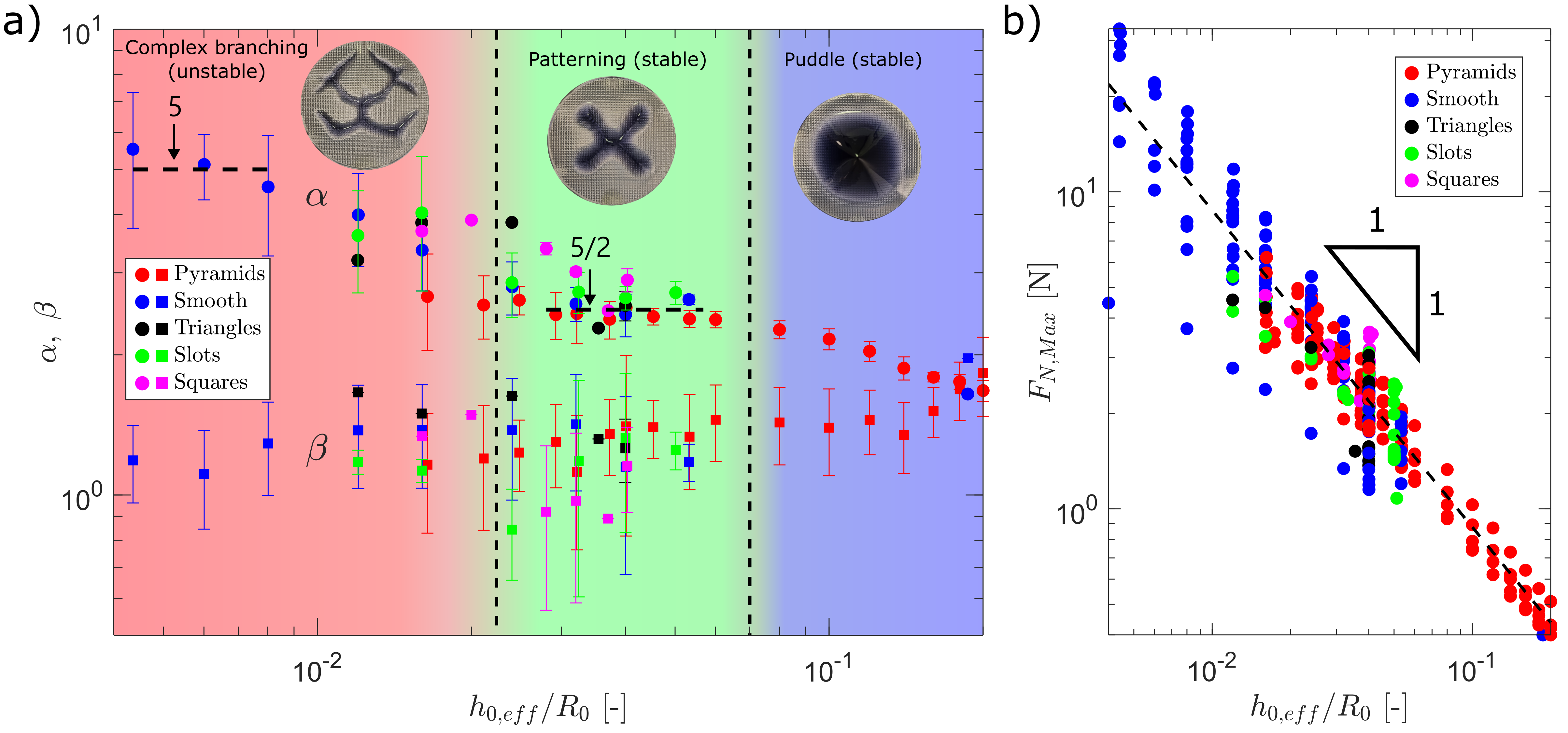}
    \caption{a) The exponents $\alpha$ (circles) and $\beta$ (squares) that describe the two scaling regimes for the force as a function of effective mean initial height $h_{0,eff}$ is extracted for the different patterns on the plate (different colors, see legend). The height is normalized by the plate radius $R_0 = 25$mm. It is notable that the patterns on the plates do not appear to influence $\alpha$ and $\beta$, where the errorbar is extracted across samples with different lifting speeds $v_l\in [10-1000]\mu$m/s. The background colors denote the approximate domains for the unstable complex branching (red), stable patterning (green) and stable puddle formation (blue), with one typical example of final pattern deposition shown for each domain. b) Maximal force as function of the effective mean initial height $h_{0,eff}$. We see that that the data follows along a scaling law $F_N\sim h_{0,eff}^{-1}$, where $h_{0,eff}$ is the mean initial film height based on the total entrapped fluid volume.}
  \label{fig:Fig4}
\end{figure}

We further characterize the evolution of the front by measuring the standard deviation of the radius $S_R$ introduced in \cite{dufresne2023} and defined as the relative deviation from the circular pattern, i.e.,

\begin{equation}
    S_R = \sqrt{ \left\langle \left( \frac{R(\zeta)}{\overline{R}} - 1 \right)^2 \right\rangle } \:,
    \label{sr}
\end{equation}
where $R(\zeta)$ is the distance to the center of the cell along the perimeter coordinate $\zeta$ and $\overline{R}$ is its average. The measurement of $S_R$ is shown by the solid brown line in Fig. 3. We notice that, as expected, this quantity stabilizes at the position marked as the third panel, also corresponding to the point where the contact line appears to pin, as seen from the side view images. This point correlates well with the approximate region of the transition between the $\alpha$ and $\beta$ scaling regimes for the force. Our results point to the fact that the pinning of the interface marks the beginning of the extensional flow part of the dynamics with the thinning of the neck connecting the upper and lower fluid parts and the $\alpha$ to $\beta$ scaling transition.

Since surface patterning can clearly affect the dynamics of the Carbopol mixture, we investigated how it affects $\alpha$, $\beta$ and the maximal normal adhesive force $F_{N,Max}$. In order to compare experiments with different surface patterns, given that a portion $V^*$ of the fluid volume is left trapped in the grooves caused by the surface patterning, we incorporate a height correction factor $\Delta = V^*/\pi R_0^2$ in the calculation of the effective mean initial height, defined as $h_{0,eff}=h_0+\Delta$. The factor $\Delta$ can be analytically determined for the surface geometries illustrated in Fig.~1d, yielding values of $\Delta = [0.25, 0.375, 0.333, 0.219]$mm for slots, squares, pyramids, and triangle patterns, respectively. In addition to the four patterns shown in Fig.~1d, we also analyzed three types of smooth geometries (a transparent Plexiglas plate, a metallic plate, and one formed by covering the surfaces with 800-grit waterproof sandpaper) as well as a beveled slot geometry (similar to the parallel slot configuration but with angled slot walls instead of parallel ones). As can be observed in Fig. 4, the slopes of the force curves $\alpha,\beta$ for different effective mean initial height $h_{0,eff}$ show that there is little difference between the textured plates despite the fact that the footprint formed by the yield stress fluid can be quite different. The background color in this figure gives a visual representation of the approximate domains for unstable complex branching (red), stable patterning (green) and stable puddle formation (blue). The complex branching seen for very small gaps $h_{0,eff}$ is characterized by a more random-looking deposition pattern, which has also been observed in other works \cite{Irm020,Barral,derks2003}. In an intermediate gap range, approximately $0.022<h_{0,eff}/R_0<0.07$ (highlighted by the vertical dashed lines), the system enters the patterning regime, where the final fluid deposition is strongly influenced by the surface geometry of the plates, (controlled deposition). For larger values of $h_{0,eff}/R_0$, we find the regime where a stable puddle is formed, also observed in previous works \cite{Irm020}. For small values of $h_{0,eff}$, we recover the expected Newtonian limit with $\alpha=5$ while at intermediate $h_{0,eff}$, the system transitions to the yield-stress-dominated regime, where $\alpha=5/2$, consistent with previous studies \cite{derks2003,Irm020}. These exponent values are indicated in Fig.~4a. For very large $h_{0,eff}$, however, we observe a deviation from the $\alpha=5/2$ scaling can be attributed to a breakdown of the Hele-Shaw condition $h_{0,eff}/R_0\ll1$ (which ensures that the flow is quasi-2D) employed in the derivation of the scaling law \cite{derks2003}.

\begin{figure}[b!]
    \includegraphics[width=1.0\columnwidth]{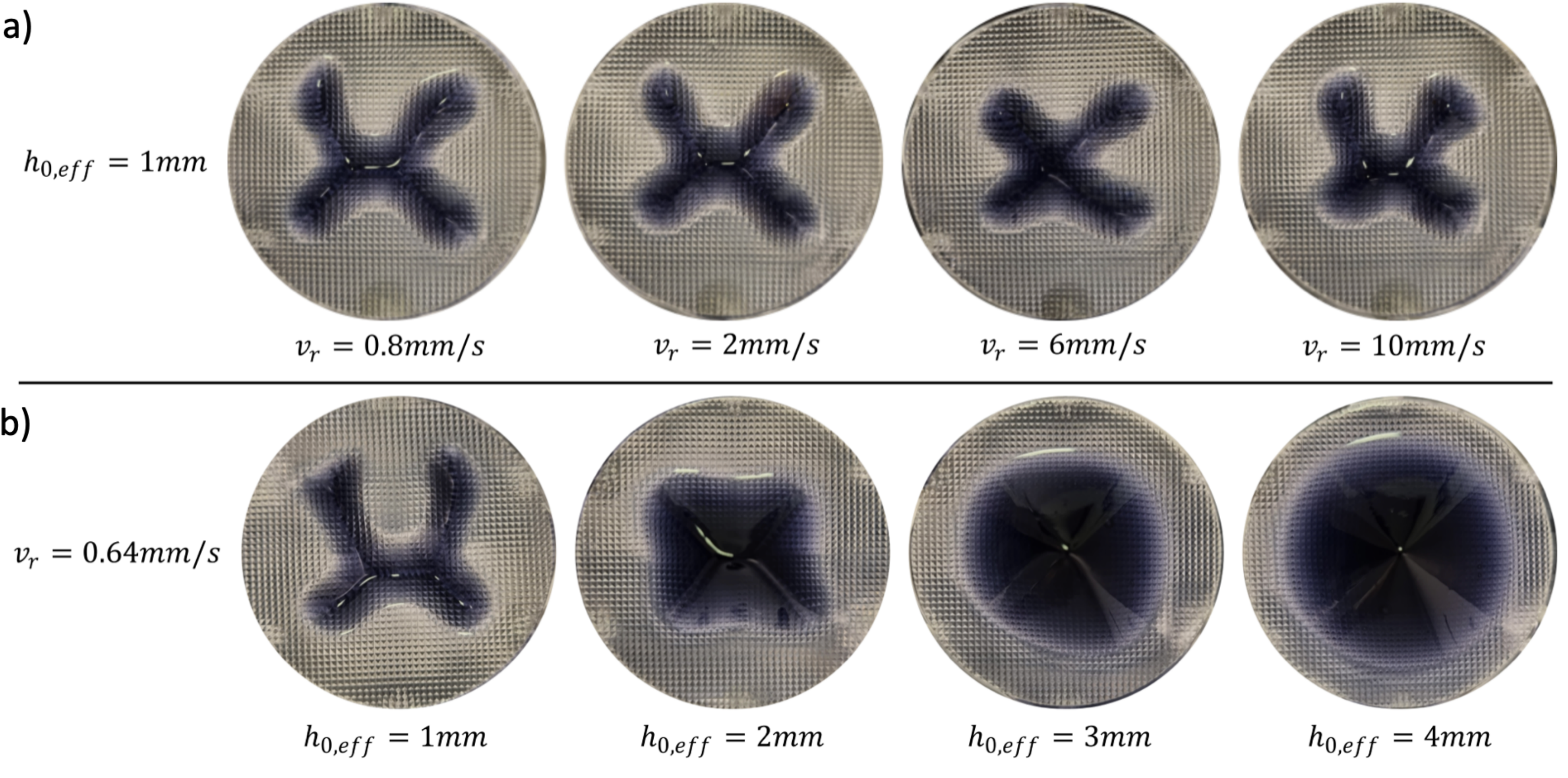}
    \caption{Sensitivity of the invasion patterns with the mean radial velocity $v_r$ and effective mean initial height $h_{0,eff}$ with $\tau=30$Pa. While the patterns seem to be fairly insensitive to $v_r$ a), they are clearly affected by $h_{0,eff}$ b), with the substrate influence on the deposition pattern only being observed for small enough $h_{0,eff}$ (corresponding to the green region in Fig. 4).}
  \label{fig:SIFig2}
\end{figure} 

Intuitively one might think that the surface pattern affects the force as more surface area is wetted in the plates' surfaces. However, by plotting the maximal normal force $F_{N,Max}$ against the effective mean initial height $h_{0,eff}$ we notice in Fig.4b  that this force appears to be unaffected. The important parameter is, as such, the trapped fluid volume (or the effective mean initial height) and not the surface structure. The observation that the maximum force $F_{N,max}$ is nearly unaffected by the surface structure can be attributed to the fact that this force is reached at a very early point in the dynamics, while the invading front is still nearly circular. This occurs well before any instabilities induced by the surface structure can develop. The maximal force also takes place at small gap sizes where the capillary force is largest and predicts a force also inversely proportional with the height $\sim (\gamma/h)R^2$.

To test the sensitivity of the lift velocity $v_l$ and the effective mean initial height $h_{0,eff}$ we systematically vary these in the experiments, see Fig. 5. When comparing between experiments with different mean film heights $h_{0,eff}$, it is instructive to control the initial radial velocity $v_r = v_l(R_0 /2 h_{0,eff})$ instead of the lift velocity $v_l$, as the same $v_l$ leads to different radial velocities depending on how much fluid is trapped, i.e., depending on $h_{0,eff}$. It appears that the final deposited pattern of yield stress fluid is fairly insensitive to the lift velocity, although it is varied more than one order of magnitude (Fig.~5a). The deposited pattern is on the other hand sensitive to the effective mean initial height $h_{0,eff}$ (Fig.~5b), in concordance with the observation in Fig.~4a where $h_{0,eff}$ affects $\alpha$ and $\beta$.

\begin{figure}[b!]
    \includegraphics[width=1.0\columnwidth]{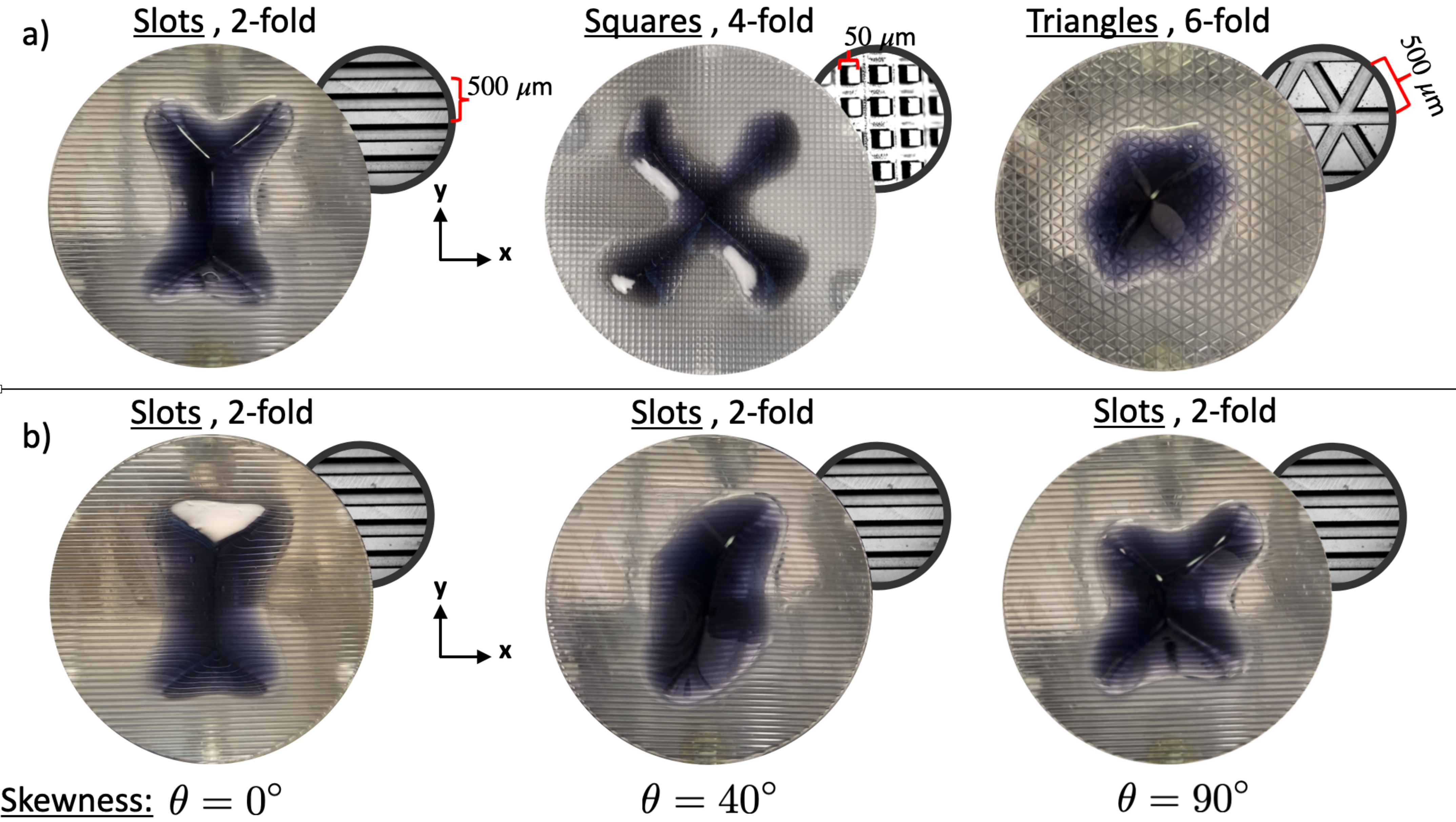}
    \caption{Illustration of how the deposited pattern of yield stress fluid can be predetermined by the structure of the pattern on the plate. a) An example of how a 2-fold (slots), 4-fold (squares) and 6-fold (triangles) pattern of the yield stress fluid can be deposited. The effective mean initial height and lifting speed were $h_{0,eff}=1$mm, $v_l=51\mu$m/s. b) Effect of plate skewness on deposition patterns. The angle $\theta$ between the slotted patterns of the upper and lower plates is varied to control their alignment. At perfect alignment ($\theta=0$), a symmetric two-fold footprint emerges. As $\theta$ increases, asymmetry develops due to differential invasion speeds along azimuthal directions. This demonstrates that pattern skewness significantly influences fluid deposition. The effective mean initial height and lifting speed were $h_{0,eff}=1.25$mm, $v_l=64\mu$m/s. See Supplemental Material for video of the evolving patterns.}
  \label{fig:Fig5}
\end{figure}

The experiments reveal that there is little effect on the force if accounting for the total entrapped volume in the gap. It is on the other hand clear that the substrate patterning can affect the final deposition of the yield stress fluid. To further illustrate this point, we explore the final deposition patterns using plates representing parallel slots, squares and triangles, see Fig. 6a. It is clear that the substrate pattern, although significantly smaller in length scale  $\sim 500$ $\mu$m to the size of the deposited footprint $\sim 10$ mm propagate upscale and affect the shape at millimeter scales. As we move from parallel slots to square to triangle shaped pattern we notice the formation of a two fold, four fold and six fold shape of the deposited yield stress fluid. It appears as if the Carbopol mixture senses the local resistance or perhaps an effective slip that is anisotropic as it contracts radially. We use the center of the plate as a point of origin/reference with the azimuthal angle starting with $\phi=0$ at the “north pole” and $\phi=90^{\circ}$ at the “equator”, with $\phi=45^{\circ}$ in between the two. Imagine moving radially from the origin to $R_0$ along the substrate at these three angles, where some different characteristics of the pattern emerge. Following along these arguments on the plates with slots, it is clear that there is a maximal number of crossings of the pattern (slots) and with a maximal mean pattern height along the radial line for $\phi=0$. While at “equator” there are no crossings and a minimal pattern height. Thus, the fluid is expected to experience a larger resistance to motion along $\phi=0$ than along the radial line of $\phi=90^{\circ}$, as also seen in Fig. 6a. Following the same lines for a pattern of squares, where the height and number of crossings are both minimal along  $\phi=0$ and $\phi=90^{\circ}$, where a clearer 4-fold symmetry is formed on the plate. For the plate patterned with triangles, we notice that the radial lines with minimal height and crossings are at $\phi=0$ and $\phi=60^{\circ}$ and the maximal height and number of crossings are along $\phi=30^{\circ}$  and $\phi=90^{\circ}$. When inspecting the deposited 6-fold pattern on the plate with triangles, the parts with largest radial extent is along $\phi\approx30^{\circ}$  and $\phi\approx90^{\circ}$, and smallest radial extent along the lines $\phi=0$ and $\phi=60^{\circ}$, following the arguments above. A video showing the evolution of the patterns for different plate geometries can be found in the Supplemental Material.

It is not only the surface pattern that can affect the footprint of the Carbopol-water mixture, but also misalignment in the pattern at the upper and lower plate in the rheometer. In Fig. 6b we illustrate this point by systematically changing the skewness of two plates, i.e., the angle $\theta$ set between the parallel slots of the upper and lower plate. If the plates are perfectly aligned $\theta=0^{\circ}$, a two-fold symmetry is formed by the deposited footprint. As the upper and lower pattern on the plates are misaligned the invasion front is moving with different speeds for the same azimuthal angle, which leads to a coupling between them through the fluid flux, particularly as the experiments enter into the regime where the neck is thinning (as the force scaling transitions from $\alpha$ to $\beta$, see Fig. 3). We then conclude that the skewness of the plates could also profoundly affect the deposition of the yield stress fluid. A video showing the evolution of the patterns for different plate skewness between top and bottom slotted patterns can be found in the Supplemental Material.

\section{Conclusion}
Controlled deposition of a liquid on a substrate is topical to a wide range of applications, where we have demonstrated that such a process involving a yield stress fluid can be controlled by macroscopic patterning of the substrate. Although substrate patterning significantly affects the dynamics of the intrusion front and the final deposited footprint of liquid, it does not affect the maximal force the plates experience as they are being moved apart. The dynamics is on the other hand sensitive to the total volume of trapped liquid (quantified through an effective mean initial height $h_{0,eff}$), where a small $h_{0,eff}$ recovers the scaling for the force from a Newtonian fluid $\sim h_{eff}^{-5}$ and for intermediate $h_{0,eff}$ the yield stress limit $\sim h_{eff}^{-5/2}$ \cite{Irm020}. We have shown that controlled deposition of the fluid (patterning) can be achieved in a range of $h_{0,eff}$, shown in green in Fig. 4a. For values of $h_{0,eff}$ outside this domain, the system either presents a random fingering pattern (small $h_{0,eff}$) or a stable cirucular puddle deposition (large $h_{0,eff}$). We hope these experiments can help inspire numerical studies of this complex interfacial dynamics, which could help identify the yielded and un-yielded portions of the flow as well as potential slip lines. Finally, these results may inspire the use of substrate patterning for the controlled deposition of yield stress fluids.

\section*{Acknowledgments}
The research has been supported by the Research Council of Norway through its projects 301138 (NANO2021 program), 262644 (PoreLab Center of Excellence) and 324555 (FlowConn Researcher Project for Young Talent).

\bibliographystyle{unsrt}

\end{document}